\documentclass[art10]{sig-alternate}

\usepackage[english]{babel}
\usepackage{amssymb,amstext,amsmath}
\usepackage{color}
\usepackage{graphicx}
\usepackage{url}
\usepackage{hyperref}
\usepackage{tikz,pgf}
\usetikzlibrary{arrows,automata,snakes,patterns}
\usepackage[sets,probability,operators]{cryptocode}
\usepackage{xspace}
\usepackage{tabularx}
\usepackage{multirow}
\usepackage{booktabs}
\usepackage{footnote}
\usepackage{paralist}
\usepackage{comment}

\newcommand{\todo}[1]{\textcolor{red}{[TODO: #1]}}

\newcommand{\erasmus}{{\small ERASMUS}\xspace}
\newcommand{\erasmusod}{{\small ERASMUS+OD}\xspace}
\newcommand{\qoa}{QoA\xspace}
\newcommand{\arm}{RM\xspace}
\newcommand{\aae}{AE\xspace}
\newcommand{\splus}{\small SMART+\xspace}
\newcommand{\smart}{\small SMART\xspace}
\newcommand{\hydra}{\small HYDRA\xspace}

\newcommand{\atp}{$Pr_{Att}$\xspace}
\newcommand{\sel}{seL4\xspace}
\newcommand{\key}{\ensuremath{K}\xspace}

\newcommand{\ptoken}{Compute $M$\xspace}
\newcommand{\measure}{\ensuremath{M}}

\newcommand{\dev}{{\ensuremath{\sf{\mathcal Prv}}}\xspace}
\newcommand{\vrf}{{\ensuremath{\sf{\mathcal Vrf}}}\xspace}
\newcommand{\prv}{{\ensuremath{\sf{\mathcal Prv}}}\xspace}

\newcommand{\highlight}[1]{\underline{#1}}

\title{{ERASMUS:} \framebox{E}fficient \framebox{R}emote \framebox{A}ttestation via 
\framebox{S}elf-\framebox{M}easurement for \framebox{U}nattended \framebox{S}ettings}

\numberofauthors{3}
\author{
\alignauthor
Xavier Carpent\\
	\affaddr{University of California, Irvine}\\
	\email{xcarpent@uci.edu}
\alignauthor
Norrathep Rattanavipanon\\
	\affaddr{University of California, Irvine}\\
	\email{nrattana@uci.edu}
\and
\alignauthor
Gene Tsudik\\
	\affaddr{University of California, Irvine}\\
	\email{gts@ics.uci.edu}
}

\begin{document}
\maketitle

\begin{abstract}
Remote attestation (RA) is a popular means of detecting malware in embedded and IoT devices. 
RA is usually realized as an interactive protocol, whereby a trusted party -- {\em verifier} -- 
measures integrity of a potentially compromised remote device -- {\em prover}. 
Early work focused on purely software-based and fully hardware-based techniques, 
neither of which is ideal for low-end devices. More recent results have yielded hybrid 
(SW/HW) security architectures comprised of a minimal set of features to support 
efficient and secure RA on low-end devices.

All prior RA techniques require \emph{on-demand operation}, i.e, RA is performed in {\em real time}. 
We identify some drawbacks of this general approach in the context of unattended devices: 
First, it fails to detect {\em mobile malware} that enters and leaves the prover between successive 
RA instances. Second, it requires the prover to engage in a potentially expensive (in terms of time 
and energy) computation, which can be harmful for critical or real-time devices.

To address these drawbacks, we introduce the concept of {\em self-measurement} where a prover device 
periodically (and securely) measures and records its own software state, based on a pre-established schedule. 
A possibly untrusted verifier occasionally collects and verifies these measurements.  We present the design of a 
concrete technique called \erasmus: {\bf E}fficient {\bf R}emote {\bf A}ttestation via {\bf S}elf-{\bf M}easurement 
for {\bf U}nattended {\bf S}ettings, 
justify its features and evaluate its performance. In the process, we also define a new metric -- \emph{Quality of 
Attestation} (\qoa). We argue that \erasmus is well-suited for time-sensitive and/or safety-critical applications that 
are not served well by on-demand RA. Finally, we show that \erasmus is a promising stepping stone towards 
handling attestation of multiple devices (i.e., a group or swarm) with high mobility.
\end{abstract}

\section{Introduction} \label{intro}
In recent years, embedded and cyber-physical systems (CPS), under the guise of {\em Internet-of-Things (IoT)},
have entered many aspects of daily life, such as: homes, office buildings, public venues, factories and vehicles. 
This trend of adding computerized components to previously analog devices and then inter-connecting them 
brings many obvious benefits. However, it also greatly expands the so-called {\em ``attack surface"} and turns 
these newly computerized gadgets into natural and attractive attack targets.  In particular, as recent incidents 
demonstrated, IoT devices can be infected with malware and used as bot-controlled
{\em zombies} in Distributed Denial-of-Service (DDoS) attacks. Also, IoT-borne malware can snoop on device
owners (by sensing) or maliciously control critical services (by actuation), as happened with Stuxnet \cite{stuxnet}.

One key component in securing IoT devices is malware detection, which is typically attained
with Remote Attestation (RA). RA is a distinct security service that allows a trusted party, 
called {\em verifier}, to securely verify the internal state (including memory and storage) of a remote
untrusted and potentially malware-infected device, called {\em prover}.
RA is realized via an interactive protocol between prover and verifier. A typical example
is described in \cite{brasser2016remote}:  (1) verifier sends an attestation request to prover,
(2) prover verifies the request\footnote{Since attestation is a potentially expensive task, this verification 
mitigates computational DoS attacks.} and (3) computes a cryptographic function of its internal state, 
then (4) sends the result to verifier, and finally, (5) verifier checks the result and decides whether prover is infected.

This general approach is referred to as \emph{on-demand attestation} and all current RA techniques adhere to it.
In this paper, we identify two important limitations of this approach. First, it is a poor match for unattended devices,
since malware that ``comes and goes'' (i.e., mobile malware~\cite{ostrovsky1991withstand}) can not be detected if it leaves prover by the time
attestation is performed. Second, for a device working under time constraints (real-time operation)
or otherwise providing critical services, on-demand attestation requires performing a possibly 
time-consuming task while deviating from the device's main function(s). 

To address these issues, we design \erasmus: {\bf E}fficient {\bf R}emote {\bf A}ttestation via {\bf S}elf-{\bf M}easurement 
for {\bf U}nattended {\bf S}ettings. \erasmus is based on \emph{self-measurements}. Basically, a device (prover) 
measures and records its state at scheduled times. Measurements are stored in prover's {\em insecure} memory. 
Verifier occasionally collects and validates these measurements in order to establish the \emph{history} of prover's state. 
Notably, with this general approach,  verifier imposes only negligible real-time burden on prover. 
It also offers strictly better quality-of-service than prior attestation techniques, because verifier obtains
prover's entire history of measurements, since the last verifier request. In other words, \erasmus de-couples 
(1) frequency of prover checking, from (2) frequency of prover measurements, which are equivalent in 
on-demand attestation. Finally, \erasmus simplifies RA design (in terms of required features) for prover: authentication of verifier requests is no longer needed, since computational DoS attacks do not arise.\footnote{This 
is unlike requirements in~\cite{brasser2016remote} that stipulate (potentially expensive) prover auhentication of verifier's requests.}

We also introduce the new notion of \emph{Quality of Attestation} (\qoa) which captures: (1) how a device (prover) is attested, 
(2) how often its state is measured, and (3) how often these measurements are verified. It is the temporal 
analogue of the concept of quality of swarm attestation (QoSA) introduced in~\cite{lisa} 
in the context of attesting groups of devices.

\noindent {\em \underline{NOTE:}}
\erasmus is not intended as a replacement for on-demand attestation, mainly because for some devices and
some settings, real-time on-demand attestation is mandatory, e.g., immediately before or after a software update or for secure erasure/reset. 
Also, on-demand attestation may be more flexible, e.g., if the verifier is only interested in  
measuring a fraction of prover's memory. These two approaches are not mutually exclusive and 
may be used together to increase \qoa, specifically, in terms of freshness of the latest measurement.

The last incentive for our self-measurement approach is its suitability for highly mobile groups of devices. 
RA protocols developed for ``swarm attestation'', e.g., \cite{seda,sana,darpa,lisa}, are designed to efficiently attest 
groups of interconnected devices on-demand, with a single verifier-prover interaction. However,
they do not work in highly mobile swarms, since on-demand attestation requires topology 
to remain essentially static during the entire attestation protocol instance -- the time for which is dominated by
computation on all swarm devices. Since \erasmus involves virtually no real time computation for prover,
it is much more suitable for high-mobility swarm settings.

After overviewing the state-of-the-art in Section~\ref{sc:ra}, we introduce \erasmus and \qoa in Section~\ref{sc:sm}. 
An implementation and experimental results are discussed in Sect~\ref{sc:exp}. Issues arising in time-sensitive 
applications and partial mitigation measures are discussed in Section~\ref{sc:time-sensitive}. Applicability of \erasmus 
to swarm attestation is considered in Section~\ref{sc:swarm}.


\section{Remote Attestation (RA)}\label{sc:ra}
%
%

RA aims to detect malware presence by \emph{verifying} integrity of a remote and untrusted 
embedded (or IoT) device. As mentioned earlier, it is typically realized as a protocol where trusted verifier 
interacts with a remote prover to obtain an integrity measurement of the latter's state.

RA techniques fall into the three main categories. (1) \emph{Hardware-based 
attestation}~\cite{Stumpf2006,SCHELLEKENS200813} uses dedicated hardware features such as a 
Trusted Platform Module (TPM) to execute attestation code in a secure environment. 
Even though such features are currently available in personal computers and smartphones, 
they are considered a relative ``luxury'' for very low-end embedded devices. 
(2) \emph{Software-based attestation}~\cite{seshadri2004swatt,Seshadri:2006:SSC:1161289.1161306} 
requires no hardware support and performs attestation solely based on precise timing measures. 
However, it limits prover to being one-hop away from verifier, so that round-trip time is either negligible or fixed. 
It also relies on strong assumptions about attacker behavior \cite{abera2016invited} and is typically only used 
for legacy devices where no other RA techniques are viable.
(3) Finally, \emph{hybrid attestation}~\cite{smart,trustlite,brasser2015tytan}, based on a software/hardware 
co-design, provides RA while minimizing its impact on underlying hardware features.

\smart\cite{smart} is the first hybrid RA design with minimal hardware modifications to existing microcontroller units (MCUs).
It has the following key features:
\begin{compactitem}
	\item Attestation code is immutable: located in and executed from ROM.
	\item Attestation code is safe: its execution always terminates
	and leaks no information other than the attestation result (token).
	\item Attestation is atomic: (1) it is uninterruptible,
	and (2) it starts from the first instruction and exits at the last instruction.
	This is realized in \smart by using hard-wired MCU access controls
	and disabling interrupts upon entering attestation code.
	\item A secret key (\key) is stored in a secure memory location where it  
	can be accessed only from within the attestation code: \key is stored in ROM and is
	guarded by specialized MCU rules.
\end{compactitem}
\cite{brasser2016remote} extended \smart to defend against denial-of-service (DoS) attacks
that try to impersonate verifier. We refers to this extended design as \splus.
\cite{brasser2016remote} additionally requires prover to have a Reliable Read-Only Clock (RROC),
which is needed to perform verifier authentication and prevent replay, reorder and delay attacks. 
To ensure reliability, RROC must not be modifiable by software.
Upon receiving a verifier request, ROM-resident attestation code checks the request's freshness using 
RROC, authenticates it, and only then proceeds to perform attestation.

The TrustLite \cite{trustlite} security architecture also supports RA for low-end devices. It differs 
from \smart in two ways: (1) interrupts are allowed and handled securely by the CPU Exception 
Engine, and (2) access control rules can be programmed using an Execution-Aware Memory 
Protection Unit (EA-MPU). TyTAN \cite{brasser2015tytan} adopts a similar approach while providing 
additional real-time guarantees and dynamic configuration for safety- and security-critical applications.

\hydra\cite{hydra} is a hybrid RA design for medium-end devices devices with a Memory Management Unit 
(MMU). It builds upon a formally verified micro-kernel, \sel\cite{klein2009sel4}, to ensure memory isolation
and enforce access control to memory regions.  Using these formally and mathematically proven features, 
access control rules can be implemented in software and enforced by \sel. 
Consequently, \hydra stores \key and attestation code in writable memory regions (e.g., flash or RAM) 
and configures the system such that no other process, besides the attestation process,
can access those memory regions. Access control configuration in \hydra also
involves the attestation process having exclusive access to its thread control block 
as well as to memory regions used for \key-related computations. 
The latter ensures the \key protection property. To ensure atomic execution, \hydra runs the attestation 
process as the {\em initial user-space process} with the highest scheduling priority, while the rest of 
user-land processes are spawned by the attestation process, with lower priorities.
Finally, hardware-enforced secure boot is used to provide integrity of \sel and the attestation process at 
system initialization time.

In this paper, we use \splus and \hydra as the base security architecture for \erasmus. 
However, \erasmus{} should be equaly applicable to other on-demand RA techniques, such as TrustLite\cite{trustlite} or TyTan\cite{brasser2015tytan}.


\section{Self-Measurements}\label{sc:sm}
As discussed in Section \ref{intro}, all current RA techniques perform \emph{on-demand attestation},
whereby prover computes verifier-requested measurements in real-time. This can be a time-consuming activity
that takes prover away from its primary mission. However, prover performs no RA-related computation
between verifier's requests.

%
%

In contrast, \erasmus divides RA into two phases. In the \emph{measurement} phase, prover 
performs self-measurements based on a pre-established schedule and stores the results. In the collection
phase, verifier (whenever it chooses to do so) contacts prover to fetch these measurements. The collection
phase is very fast since it requires practically no computation by prover. In particular, since measurements
are based on a MAC computed with a key shared between prover and verifier, no extra protection is needed
when prover sends these measurements to verifier. Furthermore, unlike in on-demand RA, there is no 
threat of computational DoS on prover. Thus, there is no need to authenticate verifier's requests, 
in contrast with on-demand attestation. 

A prover's measurement $M_t$ computed at time $t$ is defined as:
\begin{equation*}
	M_t = <t, H(mem_t), \text{MAC}_K(t, H(mem_t))>
\end{equation*}
where $H$ is a suitable cryptographic hash function and $mem_t$ represents prover's memory at time $t$. 
The computation of $H(mem_t)$ and $MAC$ is done in the context of the security architecture, e.g., \smart or \hydra.

From here on, \vrf and \dev are used to denote verifier and prover, respectively. 
Although \erasmus assumes a symmetric key $K$ shared between \vrf and \dev,
a public key signature scheme could be used instead, with no real impact on security 
of the scheme except for higher cost of measurements.


\subsection{Quality of Attestation \label{qoa}}
{\em Quality of Attestation (\qoa)} is primarily determined by two parameters: (1) time $T_M$ between two successive 
\emph{measurements} on \dev, and (2) time $T_C$ between two successive requests by \vrf 
to collect measurements from \dev. 

We assume that in most cases $T_C>T_M$.
If it so happens that $T_C\leq~T_M$, verifier will simply collect the same measurements more than once, which is redundant.
Instead, \vrf\ can {\bf explicitly request} \dev to perform a measurement before the collection.
In that case, \vrf's request would
have to be authenticated and checked for freshness (as in \splus~\cite{brasser2016remote})
before the on-demand measurement is computed. These activities clearly incur additional 
real-time overhead and delays. 
We refer to this 
variant as \erasmusod.

Exactly how $T_C$ and $T_M$ are determined clearly depends on specifics of \prv's mission 
and its deployment setting. Security impact of these parameters is intuitive. 
Smaller $T_M$ implies smaller window of opportunity for mobile malware to escape detection. 
Smaller $T_C$ implies faster malware detection. If either value is large, attestation becomes ineffective. 
Meanwhile, though low values increase \qoa, they also increase \dev's overall burden, in terms of 
computation, power consumption and communication.

Without loss of generality, we assume that measurements and collections occur at regular intervals. 
Of course, in practice this might not work in scenarios that involve critical or time-sensitive applications (see Section~\ref{sc:time-sensitive}). In fact, it might be advantageous to take measurements at irregular intervals,
as doing so might give prover a bit of an extra edge against mobile malware (see Section~\ref{sc:irreg-intervals}).

Another \erasmus parameter is the number of measurements (referred to as $k$) obtained by \vrf 
in each collection phase. It can range between {\em one} (only the most recent measurement) and {\em all}. 
In a typical setting, \prv's history size should be set such that each measurement is collected exactly once. That is, $k = \left\lceil T_C / T_M \right\rceil$. 

Finally, the collection phase involves the notion of \emph{freshness}, i.e., how recent is \dev's latest measurement. 
Depending on the application, maximal freshness might be required, e.g., right before or after a software
update. Maximal freshness is attainable via on-demand attestation
In \erasmus, freshness of a measurement (denoted as $f$) ranges between $T_M$ and $0$, which correspond
to minimal and maximal freshness, respectively. On average, we expect $f=T_M/2$.

\begin{figure}
	\begin{center}
		\begin{tikzpicture}
			\node at (.5,.25) {\dots{}};
			\node at (7.5,.25) {\dots{}};

			\draw[->,>=stealth'] (5,-1.25) -- (7,-1.25) node[midway,below] {time};

			\node[anchor=west] (legend1) at (1,-1) {measurement};
			\draw (.5,-1) -- (legend1);
			\node[anchor=west] (legend2) at (1,-1.5) {collection};
			\draw[very thick] (.5,-1.5) -- (legend2);

			\filldraw[fill=gray] (2,0) [snake=zigzag,segment amplitude=1pt,segment length=3pt] -- (2,.5) [snake=none] -- (2.3,.5) [snake=zigzag,segment amplitude=1pt,segment length=3pt] -- (2.3,0) [snake=none] -- (2,0);
			\filldraw[fill=gray] (4.5,0) [snake=zigzag,segment amplitude=1pt,segment length=3pt] -- (4.5,.5) [snake=none] -- (7,.5) -- (7,0) -- (4.5,0);
			\shade[left color=gray,right color=white] (6.25,0) -- (6.25,.5) -- (7,.5) -- (7,0) -- (6.25,0);

			\draw (0,0) -- (8,0);
			\draw (0,.5) -- (8,.5);

			\foreach \tm in {1,1.75,...,7} {\draw (\tm,-0.1) -- (\tm,0.8);}
			\draw (2.5,-0.3) -- (2.5,-.1);
			\draw[<->,>=stealth'] (1,.7) -- (1.75,.7) node[midway,above] {$T_M$};

			\foreach \tm in {3,6} {\draw[very thick] (\tm,-0.3) -- (\tm,0.6);}
			\draw[<->,>=stealth'] (3,-.2) -- (6,-.2) node[midway,below] {$T_C$};

			\draw[<->,>=stealth'] (2.5,-.2) -- (3,-.2) node[midway,below] {$f$};

			\node (comment1) at (2.15,1.5) {infection 1 (undetected)};
			\draw[->,>=stealth'] (2.15,.25) -- (comment1);
			\draw[fill=black] (2.15,.25) circle (1pt);

			\node (comment2) at (5.125,2) {infection 2 (detected)};
			\draw[->,>=stealth'] (5.125,.25) -- (comment2);
			\draw[fill=black] (5.125,.25) circle (1pt);
		\end{tikzpicture}
	\end{center}
	\caption{\qoa illustration: Infection 1 by mobile malware is undetected; Infection 2 is detected. $T_M$~is the time between two measurements, $T_C$~is the time between two collections, and $f$~is the freshness of each measurement.}
	\label{fig:qoa}
\end{figure}
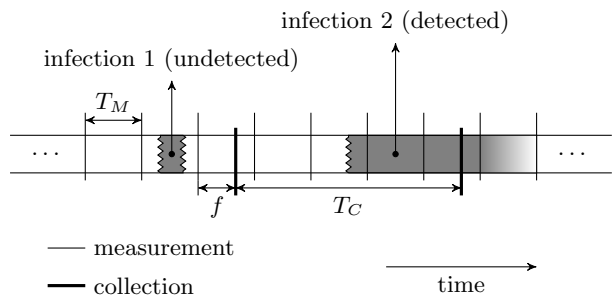

Figure~\ref{fig:qoa} shows an example with two malware infections. In the first, malware covers its tracks and 
leaves before any measurement takes place. In the second, malware persists on \dev. Although measurement 
occurs perhaps soon after infection, corrective action can be taken only after collection, thus illustrating the
importance of a small $T_C$. Measurements and collections are shown as punctual events in Figure~\ref{fig:qoa}. 
Although they do take some time to complete (measurements, in particular), it is considered negligible even for low-end
devices (see Section~\ref{sc:exp}).


\subsection{Measurements Storage \& Collection}\label{sc:rm}
A na\"ive way for \dev to store measurements is to keep track of them indefinitely. 
However, this will eventually consume a lot of \dev's storage.
To this end, \erasmus uses \emph{rolling measurements}. A fixed section of \dev's 
insecure storage is allocated as a windowed (circular) buffer for $n$ measurements. The $i$-th measurement 
is stored at location $L_{i \bmod{n}}$. However, it is expected that  \vrf collects measurements 
sufficiently often, such that no measurement is over-written. That is, the time between successive
collections should be at most $T_C \leq n \cdot{} T_M$.

The interaction between \dev and \vrf is very simple: \vrf asks for $k$ latest measurements, which \dev 
simply reads from the buffer and transmits. The collection phase does not involve any change of state 
on \dev and sent measurements are not encrypted. (Though recall that they are authenticated, since
each measurement is computed using $K$). It also does not trigger any significant computation on \dev,
i.e., in contrast with on-demand attestation, no cryptographic operations are required in
the collection phase. 
However, this is not the case in the \erasmusod variant mentioned in Section \ref{qoa}, where (1) \vrf's
request must be authenticated and checked for freshness, and (2) a current measurement must be computed.

Self-measurements can be stored in \dev's unprotected storage. This allows malware (that is possibly present on \dev)
to tamper with measurements, by modifying, re-ordering and/or deleting them. However, since malware (by design of \smart)
cannot access $K$, it cannot forge measurements. 
Thus, it is easy to see that any tampering will be detected by \vrf at the next collection phase and malware presence 
would be immediately be noticed. For the same reasons, code that handles request parsing as well as storage and transmission of measurements 
does not need to be executed in a secure environment or stored in ROM. Code that performs self-measurement, however, 
must be protected by the underlying security architecture, as in on-demand attestation.

Scheduling in \erasmus can be implemented in a very simple and stateless manner. Let $t$ be the value of RROC at the time of measurement $M_t$, and let $T_M$ be the time between two successive
measurements, as configured in \dev. The windowed buffer slot $L_i$, used to store $M_t$, is determined by:
$i = \left\lfloor t / T_M \right\rfloor \bmod n$.

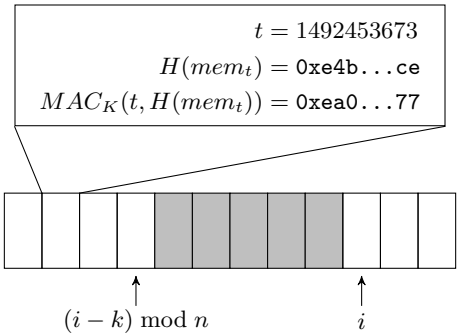
\begin{figure}
	\begin{center}
		\begin{tikzpicture}
			\foreach \x in {0,.5,...,1.5} {\draw (\x,0) rectangle (\x+.5,1);}
			\foreach \x in {2,2.5,...,4} {\draw[fill=gray!50!white] (\x,0) rectangle (\x+.5,1);}
			\foreach \x in {4.5,5,...,5.5} {\draw (\x,0) rectangle (\x+.5,1);}
			\draw[->,>=stealth'] (1.75,-.5) -- (1.75,-.1);
			\draw[->,>=stealth'] (4.75,-.5) -- (4.75,-.1);
			\node at (1.75, -.7) {$(i - k) \bmod n$};
			\node at (4.75, -.7) {$i$};
			\node[draw,text width=5.5cm,text justified] (block) at (3, 2.7) {\vspace{-2.5mm}\begin{align*}t&=1492453673\\H(mem_t)&=\texttt{0xe4b...ce}\\MAC_K(t, H(mem_t))&=\texttt{0xea0...77}\end{align*}};
			\draw (.5,1) -- (block.south west);
			\draw (1,1) -- (block.south east);
		\end{tikzpicture}
	\end{center}
	\caption{\erasmus memory allocation. Example with $n=12, i=3, k=7$.}
	\label{fig:rmmem}
\end{figure}

\erasmus collection protocol is shown in Figure~\ref{fig:rmprot}. No operation involves the underlying architecture during 
collection; only during measurements. Notation $^*\!L_j$ refers to contents of location $L_j$. A sample memory layout is shown 
in Figure~\ref{fig:rmmem}.

\begin{figure}
	\begin{center}
		\fbox{
		\procedure[codesize=\scriptsize]{}{
			\\[-2mm]
			\textbf{\normalsize\vrf} \< \< \textbf{\normalsize\dev}\\[1mm][\hline]
			\\[-2mm]
			\< \hspace{-7mm}\sendmessageright*[3.9cm]{\text{collect } k} \<\\
			\< \< \pcif k > n:\\
			\< \< \pcind k = n\\
			\< \hspace{-7mm}\sendmessageleft*[3.9cm]{M=\{^*\!L_{(i - j) \bmod n} \mid 0 \leq j < k\}} \<\\
			\pcforeach M_t \in M: \< \<\\
			\pcind \text{check } t \text{ and } h \< \<\\
			\pcind \text{verify } \text{MAC}_K(t, h) \< \<\\
		}}
	\end{center}
	\caption{\erasmus collection protocol.}
	\label{fig:rmprot}
\end{figure}

\subsection{\erasmusod: \erasmus with On-demand Attestation}\label{sc:combining}
As mentioned in Section \ref{qoa}, \erasmus may be combined with on-demand attestation to benefit from 
advantages of both approaches. This variant, \erasmusod, records \dev's state history to detect 
mobile malware, and uses on-demand attestation to obtain better freshness. Freshness is particularly 
relevant whenever real-time attestation is mandatory, e.g., immediately before or after a software update.

\begin{figure}
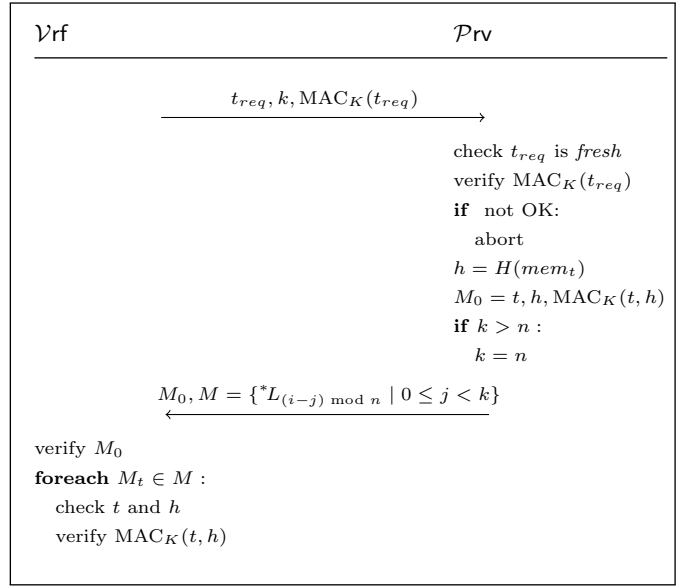

	\begin{center}
		\vspace{-2.6mm} 
		\fbox{
		\procedure[codesize=\scriptsize]{}{
			\\[-2mm]
			\textbf{\normalsize\vrf} \< \< \textbf{\normalsize\dev}\\[1mm][\hline]
			\\[-2mm]
			\< \hspace{-11mm}\sendmessageright*[4.3cm]{t_{req}, k, \text{MAC}_K(t_{req})}\hspace{-11mm} \<\\
			\< \< \text{check } t_{req} \text{ is \emph{fresh}}\\
			\< \< \text{verify } \text{MAC}_K(t_{req})\\
			\< \< \pcif \text{ not OK:}\\
			\< \< \pcind \text{abort}\\
			\< \< h = H(mem_t)\\
			\< \< M_0 = t, h, \text{MAC}_K(t, h)\\
			\< \< \pcif k > n:\\
			\< \< \pcind k = n\\
			\< \hspace{-11mm}\sendmessageleft*[4.3cm]{M_0, M=\{^*\!L_{(i - j) \bmod n} \mid 0 \leq j < k\}}\hspace{-11mm} \<\\
			\text{verify } M_0 \< \<\\
			\pcforeach M_t \in M: \< \<\\
			\pcind \text{check } t \text{ and } h \< \<\\
			\pcind \text{verify } \text{MAC}_K(t, h) \< \<\\
		}}
	\end{center}
	\caption{ERASMUS+OD protocol.}
	\label{fig:eod}
\end{figure}

The measurement phase is not modified, while the collection phase is combined with on-demand 
attestation request as follows. First, as part of each attestation request \vrf now computes and includes an 
authentication token and specifies $k$. As in \splus~\cite{brasser2016remote}, authentication of \vrf protects \dev against computational DoS. 
Then, only after checking that a request is valid, \dev computes a measurement. Finally, this real-time measurement 
is sent to \vrf, along with $k$ previous measurements. This protocol is shown in Figure \ref{fig:eod}.

This anti-DoS protection incurs an additional cost for \dev which may interfere with its normal function. A major advantage of \erasmus over \erasmusod and regular on-demand attestation is that no such protection is required.


\subsection{Security Considerations}\label{sc:security}
Security of the measurement process itself is based on the underlying security architecture, 
e.g., \splus{} or \hydra{}, which: (1) provides measurements code with exclusive access to \key, 
(2) ensures non-malleability and non-interruptibility of the measurement code, and (3) performs 
memory-cleanup after execution.

The timestamps used in the measurement process \emph{must} be based on the RROC which 
(by definition) can not be modified by non-physical means. This is important since 
malware should not  influence when measurements are taken. 

If RROC value could be modified, the following attack scenario would become possible: malware enters at time 
$t_0$ and remains active long enough so that a measurement at time $t_0 + \delta$ (with $\delta < T_M$) is taken. 
Before leaving, malware discards that measurement and resets the counter to $t_0$. 
Soon after $\delta$ (so that a measurement, valid this time, has been taken for $t_0 + \delta$), malware returns and resets 
the counter to time elapsed since $t_0$. Though this example works for one $T_M$ window, it can be extended to 
arbitrarily many. It requires an additional assumption that no collection took place during the presence of malware.

Fortunately, RROC is already a requirement of the underlying \splus security architecture, for a totally different reason. 
In \splus, RROC helps prevent replay and computational DoS attacks on \dev. Thus, \erasmus does not require any 
changes to the underlying security architecture.

As mentioned earlier, measurements need not be stored in protected memory because 
tampering with them is detectable and indicates malware presence on \dev.
Likewise, the code to support the collection phase does not require any protection 
since measurements are not secret (they are unique for every device and every timestamp value), 
and their absence or alteration is self-incriminating.


\subsection{Irregular Intervals}\label{sc:irreg-intervals}

A natural extension to \erasmus{} is to use irregular measurements intervals instead of a fixed $T_M$. The motivation is that mobile malware that is aware of fixed scheduling knows when to enter/leave the device in order to stay undetected.

One way to implement irregular intervals is to use a Cryptographically Secure Pseudo Random Number Generator (CSPRNG) iniatialized (seeded) with the secret key $K$. Output of the CSPRNG can be truncated such that $T_M$ is upper- and/or lower-bounded.

For example, after computing $M_{t_i}$, \dev can set the measurement timer to:
\[T_M^{\text{next}} = \operatorname{map}(\operatorname{CSPRNG}_k(t_i)),\]
where $\operatorname{map}$ is a function that maps CSPRNG output to seconds, e.g. $\operatorname{map} : x \mapsto x \bmod{(U-L)} + L$, with $U$ and $L$ upper and lower bounds, respectively.

The timer itself must be read-protected to ensure that $T_M^{\text{next}}$ is unknown to malware potentially present on \dev. CSPRNG code must be protected in the same way as the measurement collection.



\section{Implementation}\label{sc:exp}
%
We implemented \erasmus on two security architectures: \splus and \hydra.
The main difference between them is that the former targets low-end devices,
and the latter -- medium-end devices with a memory management unit (MMU).

\subsection{Implementation on \splus}
\begin{figure}[!t]
	\tiny
	\tikzstyle{rect}=[draw,rectangle, minimum width=2cm, minimum height=3.5mm, anchor=south, yshift=-3.5mm]
	\tikzstyle{rect2}=[draw,rectangle, minimum width=2cm, minimum height=7mm, anchor=south, yshift=-7mm]
	\tikzstyle{rect3}=[draw,rectangle, minimum width=2cm, minimum height=10.5mm, anchor=south, yshift=-10.5mm]
	\tikzstyle{rect4}=[draw,rectangle, minimum width=2cm, minimum height=14mm, anchor=south, yshift=-14mm]
	\tikzstyle{rect5}=[draw,rectangle, minimum width=2cm, minimum height=17.5mm, anchor=south, yshift=-17.5mm]
	\tikzstyle{invis_rect1}=[rectangle, minimum width=2cm, minimum height=3mm, anchor=south, yshift=-3mm]
	\tikzstyle{invis_rect}=[rectangle, minimum width=10mm, minimum height=1mm]

	\resizebox{\linewidth-3.1mm}{!}{%
	\begin{tabular}{cccc}
		\hline
		\begin{tikzpicture}
			\node[invis_rect, align=center] {RAM/\\Flash};
		\end{tikzpicture}
		&\hspace{3.6mm}
		\begin{tikzpicture}
			\node[rect3] (ram_gap) {};
			\node[rect] (token) at (ram_gap.south) {\measure};
		\end{tikzpicture}
		&\hspace{2.6mm}
		\begin{tikzpicture}
			\node[rect] (ram_gap) {};
			\node[rect] (mn) at (ram_gap.south) {\measure$_n$};
			\node[rect] (mdot) at (mn.south) {...};
			\node[rect] (m1) at (mdot.south) {\measure$_1$};
		\end{tikzpicture}
	\end{tabular}}

	\vspace{-1.1mm}

	\resizebox{\linewidth-3mm}{!}{%
	\begin{tabular}{cccc}
		\hline
		\begin{tikzpicture}
			\node[invis_rect] {ROM};
		\end{tikzpicture}
		&
		\begin{tikzpicture}
			\node[rect,fill=black!40] (t1) {};
			\node[rect,fill=black!40] (key) at (t1.south) {\key};
			\node[rect2,fill=black!40, align=center] (att) at (key.south) {AuthRequest\\\& \ptoken};
			\draw [->] (att) to [out=180,in=180,left] node {r} (key);
		\end{tikzpicture}
		&
		\begin{tikzpicture}
			\node[rect2,fill=black!40] (t1) {};
			\node[rect,fill=black!40] (key) at (t1.south) {\key};
			\node[rect,fill=black!40] (pm) at (key.south) {\ptoken};
			\draw [->] (pm) to [out=180,in=180,left] node {r} (key);
		\end{tikzpicture}
	\end{tabular}}
	\vspace{-1.5mm}

	\resizebox{\linewidth-3.1mm}{!}{%
	\begin{tabular}{cccc}
		\hline
		\begin{tikzpicture}
			\node[invis_rect] {I/O};
		\end{tikzpicture}
		&\hspace{3.6mm}
		\begin{tikzpicture}
			\node[rect] (timer) {Timer};
			\node[rect,fill=black!40] (clock) at (timer.south) {Clock};
		\end{tikzpicture}
		&\hspace{2.6mm}
		\begin{tikzpicture}
			\node[rect] (timer) {Timer};
			\node[rect,fill=black!40] (clock) at (timer.south) {Clock};
		\end{tikzpicture}
	\end{tabular}}
	\vspace{-1.1mm}

	\resizebox{\linewidth-3.1mm}{!}{%
	\begin{tabular}{cccc}
		\hline
		\begin{tikzpicture}
			\node[invis_rect] {};
		\end{tikzpicture}
		&
		\begin{tikzpicture}
			\node[invis_rect] (timer) {};
			\node[invis_rect] (t1) at (timer.south) {};
			\node[invis_rect1] (clock) at (t1.south) {\scriptsize (a) On-Demand};
		\end{tikzpicture}
		&
		\begin{tikzpicture}
			\node[invis_rect] (timer) {};
			\node[invis_rect] (t1) at (timer.south) {};
			\node[invis_rect1] (clock) at (t1.south) {\scriptsize (b) ERASMUS};
		\end{tikzpicture}
	\end{tabular}}
	\scriptsize
	\begin{center}
	\vspace{-1em}
	\tikz \draw[fill=black!40] (0,0) rectangle (.25,.25); Hardware-Enforced Read-Only
	\end{center}
	\vspace{-2em}
	\caption{Memory organization and access rules of \splus-based RA. $r$ denotes exclusive read-only access.}
	\vspace{-2em}
	\label{fig:splus}
\end{figure}
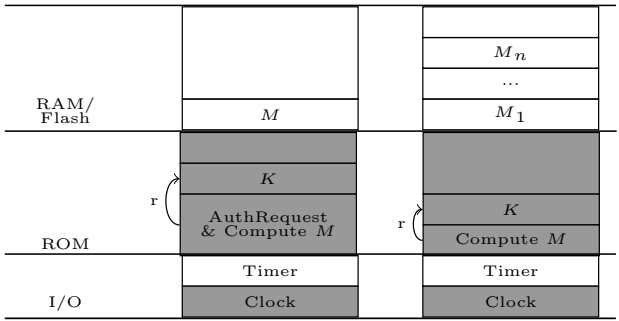

Figure~\ref{fig:splus} shows the implementation of \erasmus atop the
\splus\ architecture.
As in \splus,  measurement code and \key reside in ROM.
However, the code is invoked periodically and autonomously, whenever 
a scheduled timer interrupt occurs. 
We now examine ROM size, hardware costs and run-time on \splus architecture.

\begin{table}
\centering
\caption{Size of Attestation Executable}
\label{code-size}
  \resizebox{\linewidth}{!}{%
\begin{tabular}{lcccc}
	\toprule 
	MAC Impl. & \multicolumn{2}{c}{\splus} & \multicolumn{2}{c}{\hydra}\\
	\cline{2-5}
	& On-Demand & \erasmus & On-Demand & \erasmus\\
	\hline
    	\midrule
	HMAC-SHA1 & 4.9KB & 4.7KB & - & -\\
	HMAC-SHA256 & 5.1KB & 4.9KB & 231.96KB & 233.84KB\\
	Keyed BLAKE2S & 28.9KB & 28.7KB & 239.29KB & 241.17KB\\
    	\bottomrule
\end{tabular}%
}
\end{table}

\noindent\textbf{ROM Size} greatly depends on the choice of MAC algorithms.
We implement ROM-resident code in "C" using three MAC functions: HMAC-SHA1~\cite{sha1}\footnote{Note 
that HMAC-SHA1 is used for comparison purposes only. We exclude it in our actual implementations due to a 
recent collision attack in SHA1~\cite{sha1-atk}.}, HMAC-SHA256~\cite{sha256} and keyed 
BLAKE2S~\cite{blake2}. We then use open-source MSP430-gcc compiler~\cite{msp-gcc} to compile the 
"C" code into an MSP430 executable. Table~\ref{code-size} shows the ROM size 
for each \splus-based approach. As expected, \erasmus requires slightly less 
ROM than on-demand attestation.


\noindent\textbf{Hardware Cost:} 
We implement the hardware part of \erasmus by modifying the MSP430 architecture, using
open-source OpenMSP430 core~\cite{openmsp430}.
We modify the memory backbone module in the OpenMSP430 core to support atomic 
execution of ROM code and exclusive access to \key. RROC is realized as a peripheral 
using a 64-bit register incremented for every clock cycle. 
To ensure write-protection, a write-enable wire is removed in the RROC module.
For timer components, we use the unmodified version of omsp\_timerA module provided by OpenMSP430. 
Note that hardware timers are not considered to represent additional hardware cost.
This is because they are common and crucial components of embedded systems.
Indeed, it is unusual to find an embedded device not equipped with at least one timer.
Finally, we use Xilinx ISE 14.7~\cite{xilinx2013} to synthesize our modifications of the MSP430 core from 
a hardware description language to a combination of registers and look-up tables that serve as 
building blocks in FPGA. 

As expected, our synthesized results show that \erasmus utilizes the same amount of registers
and look-up tables as the on-demand attestation.
Compared to the unmodified MSP430 core, \erasmus requires roughly 13\% ($655$ vs. $579$) 
and 14\% ($1,969$ vs. $1,731$) 
additional registers and look-up tables respectively.

\begin{figure}[!t]
	\centering
	\includegraphics[width=\columnwidth]{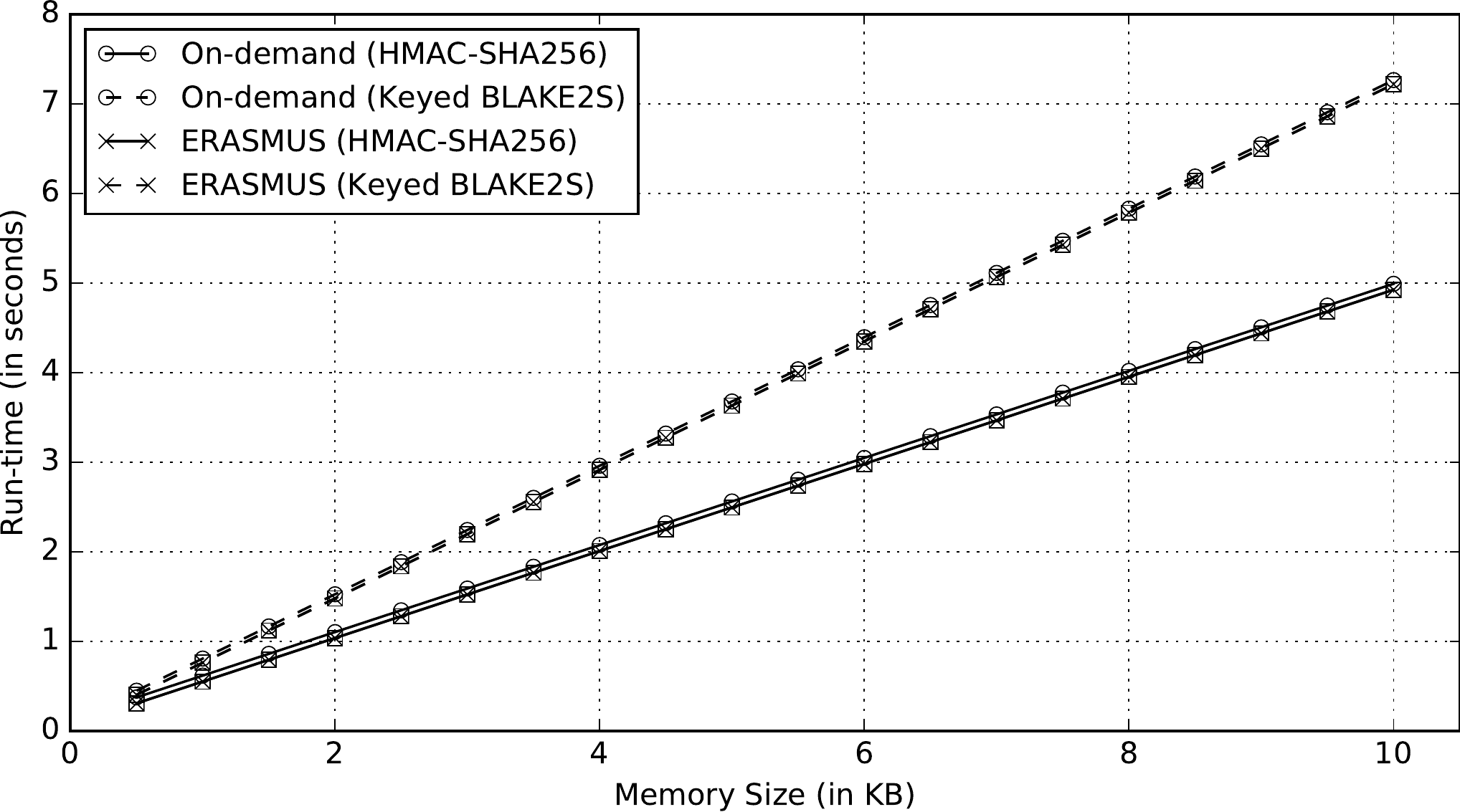}
	\vspace{-2em}
	\caption{Measurement Run-Time on MSP430-based Device @ 8MHz}
	\label{fig:splus-runtime}
\end{figure}

\noindent\textbf{Measurement Run-Time:}
Figure~\ref{fig:splus-runtime} illustrates run-time of the measurement phase for various memory sizes. 
Not surprisingly, it is linearly dependent on memory size and roughly equivalent to that 
of on-demand attestation.

\subsection{Implementation on \hydra}
\begin{figure}[!t]
	\tiny
	\tikzstyle{rect}=[draw,rectangle, minimum width=2cm, minimum height=3.5mm, anchor=south, yshift=-3.5mm]
	\tikzstyle{rect2}=[draw,rectangle, minimum width=2cm, minimum height=7mm, anchor=south, yshift=-7mm]
	\tikzstyle{rect3}=[draw,rectangle, minimum width=2cm, minimum height=10.5mm, anchor=south, yshift=-10.5mm]
	\tikzstyle{rect4}=[draw,rectangle, minimum width=2cm, minimum height=14mm, anchor=south, yshift=-14mm]
	\tikzstyle{rect5}=[draw,rectangle, minimum width=2cm, minimum height=17.5mm, anchor=south, yshift=-17.5mm]
	\tikzstyle{invis_rect1}=[rectangle, minimum width=2cm, minimum height=3mm, anchor=south, yshift=-3mm]
	\tikzstyle{invis_rect}=[rectangle, minimum width=10mm, minimum height=1mm]

	\resizebox{\linewidth-3.1mm}{!}{%
	\begin{tabular}{cccc}
		\hline
		\begin{tikzpicture}
			\node[invis_rect, align=center] {RAM/\\Flash};
		\end{tikzpicture}
		&
		\begin{tikzpicture}
			\node[rect2] (t1) {};
			\node[rect, fill=black!20] (token) at (t1.south) {\measure};
			\node[rect, fill=black!20] (key) at (token.south) {\key};
			\node[rect2, align=center, fill=black!20] (att) at (key.south) {AuthRequest\\\& \ptoken};
			\node[rect] (sel) at (att.south) {\sel Kernel};
		\end{tikzpicture}
		&
		\begin{tikzpicture}
			\node[rect] (t1) {};
			\node[rect, fill=black!20] (mn) at (t1.south) {\measure$_n$};
			\node[rect, fill=black!20] (mdot) at (mn.south) {...};
			\node[rect, fill=black!20] (m1) at (mdot.south) {\measure$_1$};
			\node[rect, fill=black!20] (key) at (m1.south) {\key};
			\node[rect, fill=black!20] (pm) at (key.south) {\ptoken};
			\node[rect] (sel) at (pm.south) {\sel Kernel};
		\end{tikzpicture}
	\end{tabular}}

	\vspace{-1.1mm}

	\resizebox{\linewidth-3.1mm}{!}{%
	\begin{tabular}{cccc}
		\hline
		\begin{tikzpicture}
			\node[invis_rect] {ROM};
		\end{tikzpicture}
		&
		\begin{tikzpicture}
			\node[rect, fill=black!40] (sb) {Secure Boot};
		\end{tikzpicture}
		&
		\begin{tikzpicture}
			\node[rect, fill=black!40] (sb) {Secure Boot};
		\end{tikzpicture}
	\end{tabular}}
	\vspace{-1.1mm}

	\resizebox{\linewidth-3.1mm}{!}{%
	\begin{tabular}{cccc}
		\hline
		\begin{tikzpicture}
			\node[invis_rect] {I/O};
		\end{tikzpicture}
		&
		\begin{tikzpicture}
			\node[rect] (timer) {Timer};
			\node[rect] (clock) at (timer.south) {Clock};
		\end{tikzpicture}
		&
		\begin{tikzpicture}
			\node[rect] (timer) {Timer};
			\node[rect] (clock) at (timer.south) {Clock};
		\end{tikzpicture}
	\end{tabular}}
	\vspace{-1.05mm}

	\resizebox{\linewidth-3.1mm}{!}{%
	\begin{tabular}{cccc}
		\hline
		\begin{tikzpicture}
			\node[invis_rect] {};
		\end{tikzpicture}
		&
		\begin{tikzpicture}
			\node[invis_rect] (timer) {};
			\node[invis_rect] (t1) at (timer.south) {};
			\node[invis_rect1] (clock) at (t1.south) {\scriptsize (a) On-Demand};
		\end{tikzpicture}
		&
		\begin{tikzpicture}
			\node[invis_rect] (timer) {};
			\node[invis_rect] (t1) at (timer.south) {};
			\node[invis_rect1] (clock) at (t1.south) {\scriptsize (b) ERASMUS};
		\end{tikzpicture}
	\end{tabular}}
	
	\scriptsize
	\begin{center}
	\vspace{-1em}
	\tikz \draw[fill=black!40] (0,0) rectangle (.25,.25); Hardware-Enforced Read-Only
	\hspace{5mm}
	\tikz \draw[fill=black!20] (0,0) rectangle (.25,.25); Attestation Process
	\vspace{-2em}
	\end{center}

	\caption{Memory organization of \hydra-based on-demand attestation and \erasmus.}
	\vspace{-3em}
	\label{fig:hydra}
\end{figure}

Figure~\ref{fig:hydra} illustrates implementations of \hydra-based \erasmus and
on-demand attestation. 
We implement these two techniques on an I.MX6 Sabre Lite~\cite{sabre-lite} development board.
RROC is implemented based on the software clock approach, suggested by Brasser et al.~\cite{brasser2016remote}.
Specifically, we use a short-term counter from Sabre Lite's General Purpose Timer (GPT) and our 
clock code in \atp to construct RROC. When the counter wraps around and causes an interrupt, 
our clock code handles it by updating higher-order bits of the clock in \atp. Then, the clock value is 
constructed by combining these bits with the GPT counter.
To ensure read-only property, \atp is given exclusive write-access to RROC components.
Also, we utilize Sabre Lite's Enhanced Periodic Interrupt Timer (EPIT) to schedule
execution of \erasmus measurement code

We base the code of \atp on open-source \sel libraries~\cite{sel4-libs}: seL4utils, seL4vka, seL4vspace,
and seL4bench. The first three provide abstractions of: process, memory management 
and virtual space, respectively, while the last one is used to evaluate performance. 
Finally, we use~\cite{sel4-util-libs} to implement th network stack: an Ethernet driver and timer drivers in \sel.

\noindent\textbf{Executable Size:}
Table~\ref{code-size} compares executable sizes of \atp in on-demand attestation and \erasmus.
Results show that \erasmus is only about 1\% higher in terms of the executable size.
This overhead mostly comes from the need for an additional timer driver.

\begin{figure}[!t]
	\centering
	\includegraphics[width=\columnwidth]{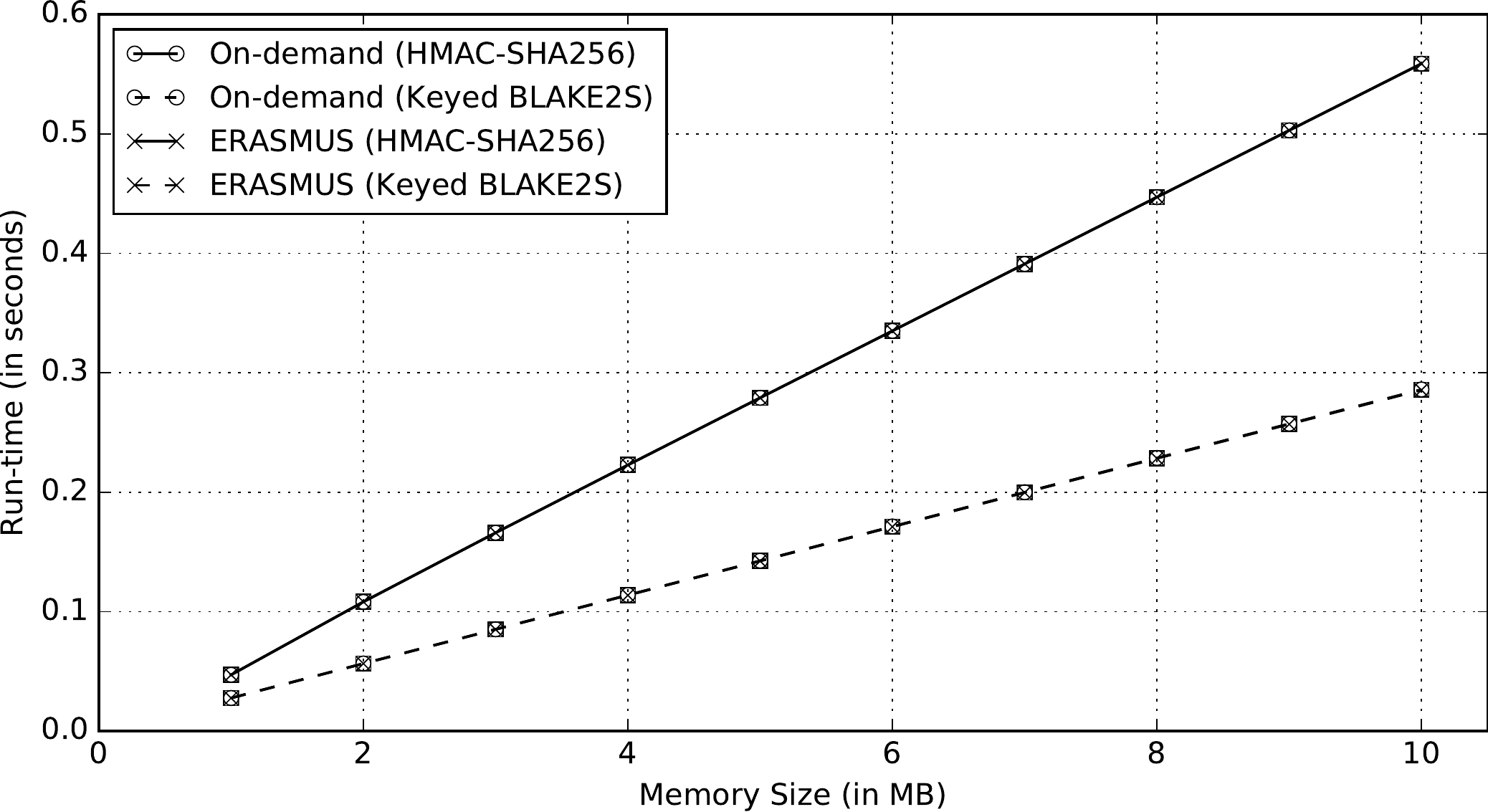}	
	\label{fig:hydra-runtime}
	\vspace*{-0.7cm}
	\caption{Measurement Run-Time on I.MX6 Sabre Lite @ 1GHz}
\end{figure}

\noindent\textbf{Measurement Run-time:}
Measurement run-time of \hydra-based \erasmus in Figure~\ref{fig:hydra-runtime} follows the same trend 
as \splus-based \erasmus: (1) it is linear as a function
of memory sizes, and (2) it is roughly equal to that of on-demand attestation.

\noindent\textbf{Collection Run-time:}
Table~\ref{fig:hydra-collect} shows the run-time breakdown of the collection phase
for each variant.
Clearly, in \erasmus, 
run-time of the collection phase is negligible (by at least a factor of $3,000$), 
compared to that of the measurement phase.
Collection run-time in ERASMUS+OD, on the other hand, is dominated by run-time
of performing on-demand attestation.

%


\begin{savenotes}
\begin{table}
\centering
\label{fig:hydra-collect}\caption{Run-Time (in ms) of Collection Phase on I.MX6-Sabre Lite}
\resizebox{.8\columnwidth}{!}{%
\begin{tabular}{lcc}
	\toprule 
	Operations & \erasmus & ERASMUS+OD\\
	\hline
    	\midrule
	Verify Request & N/A & 0.005\\
	Compute Measurement\footnote{On 10MB memory using keyed BLAKE2S as the underlying MAC function.} & N/A & 285.6\\
	Construct UDP Packet & 0.003 & 0.003\\
	Send UDP Packet & 0.012 &  0.012\\
	\hline\\[-3mm]
	Total Collection Run-time & 0.015 & 285.6\\
    	\bottomrule
\end{tabular}	
}
\end{table}
\end{savenotes}

\section{Availability in Time-Sensitive Applications}\label{sc:time-sensitive}
In some cases, it might be undesirable to interrupt execution of the \prv's application process
in order to obtain a measurement. This is particularly the case for time-sensitive or safety-critical 
applications. As discussed in Section \ref{sc:exp}, measurements can take non-negligible time, e.g., 
$7$ seconds on an 8-MHz device with 10KB RAM. Making \prv unavailable for that long is 
not appropriate.

As is, pure on-demand attestation is poorly suited for such applications. At the same time, if \prv follows a strict schedule, 
\erasmus is also not a remedy since it suffers from the same issue. However, it can be made more flexible.

One partial measure is for \dev to be self-aware of when time-sensitive tasks occur. That way, it can schedule 
measurements at appropriate times. If this knowledge is also available to \vrf, on-demand attestation could be 
used if \vrf adapts to \prv's schedule. 

Another approach is to allow \prv to abort the measurement in progress, if the need arises. However, this has some
caveats: First, the security architecture needs to be adapted to allow interrupts during measurements. Protection of keys 
(and cleanup in case of an interrupt) is still required; thus, there is still a need for some hardware support. Second, 
it would be trivial for malware to abort computation of measurements in order to avoid detection, or simply pretend,
when queried by \vrf, that all attempted measurements have been aborted. Therefore, \vrf must use some external 
information or policy to decide whether there is a valid justification for each aborted measurement.

To handle such situations, we consider another \erasmus variant that involves \emph{lenient scheduling}. 
Instead of performing a measurement every $T_M$, \dev has a window of $w\times~T_M$ where $w\geq~1$. 
Under normal conditions, \dev behaves as usual, using the $T_M$ window. If something causes a measurement 
to be aborted, it can be rescheduled to the end of the current window.

These are certainly not ideal measures, the underlying problem seems quite difficult to address deterministically.
As is typical for security/usability compromises, real deployment would likely involve policy-based decisions. 


\section{Swarm Attestation}\label{sc:swarm}
Some applications require attesting a group (or swarm) of interconnected embedded devices. In such a setting, 
it is beneficial to take advantage of interconnectivity and perform collective attestation using a dedicated protocol. 
Several swarm attestation techniques have been proposed. SEDA~\cite{seda} is the first such scheme, which
relies on hybrid attestation security architectures: \smart~\cite{smart} and TrustLite~\cite{trustlite}. SEDA 
combines them with a request-flooding and response-gathering protocol. SEDA was improved and further 
specified in LISA~\cite{lisa}. Other related techniques deal with report aggregation~\cite{sana} or physical attacks~\cite{darpa}.

A concept of Quality of Swarm Attestation (QoSA) was introduced in~\cite{lisa} to capture  the level of information 
that \vrf\ obtains as a result of swarm attestation. This can range from binary (``is the whole swarm healthy?'') to full 
(state of each individual device and topology information). \qoa, as introduced in this paper, is an orthogonal measure 
that captures the state of a given device in time. \qoa and QoSA can be used in concert with one another.

\erasmus could be used instead of on-demand attestation in the context of swarm RA protocols. In particular, 
\prv self-measurements can be coupled with a collection protocol, such as LISA-$\alpha$, where the latter 
only relays reports and does not perform any computation. This would yield a clean and conceptually simple approach
to swarm attestation, with all the benefits of \erasmus.

An additional advantage of using \erasmus in the swarm setting is support for high mobility. Prior swarm RA techniques, such 
as SEDA, SANA and LISA require swarm topology to remain almost static during the whole swarm attestation instance. 
This process may be long and prohibitive for applications where connectivity changes often. 
\erasmus does not require external input and its collection phase is very fast, since it does not involve 
any computation; only reading and sending stored measurements. This makes \erasmus a very natural and 
viable technique for highly-mobile swarms.

Finally, related to the discussion in Section~\ref{sc:time-sensitive}, we consider the scenario where availability of at least 
one in (or a part of) a group of devices is required at all times. This cannot be guaranteed by on-demand swarm attestation, 
where a large part of the network may be concurrently busy. Meanwhile, with \erasmus, it is trivial to establish a 
schedule which ensures that only a fraction of the swarm computes measurements at any given time.


\section{Conclusion}
We designed \erasmus as an alternative to current methods that perform 
on-demand RA for low-end devices. \erasmus provides better \qoa in that it
allows \vrf to detect mobile malware, which is not possible with on-demand techniques
that only detect malware if it is currently on \prv. \erasmus makes it harder for malware to avoid 
detection. \erasmus's other major advantage is that it requires no cryptographic computation by \prv
as part of its interaction with \vrf. This is particularly relevant in time-sensitive and critical 
applications, where \prv's availability is very important. We discuss partial mitigation measures for this
problem.

We present the new notion of Quality-of-Attestation (\qoa) as a measure of temporal security 
guarantees given by an attestation technique. We show that timing of measurements and timing 
of verifications (that are conjoined in on-demand attestation) are two distinct aspects of \qoa. 
They are treated as distinct parameters in \erasmus. We also discuss that the possibility of 
using on-demand attestation as part of \erasmus collection phase to obtain maximal
freshness.

We implemented \erasmus on two hybrid RA architectures, \splus and \hydra, and demonstrated  
its viability on both. \erasmus does not require extra features or a larger ROM than what is 
needed in \splus, and each measurement is fast than on-demand attestation since no authentication
of \vrf requests is needed. Finally, we show that \erasmus is a promising option for highly-mobile 
groups/swarms of devices, for which no current RA technique works well.



\begin{thebibliography}{10}

\bibitem{abera2016invited}
Tigist Abera, N~Asokan, Lucas Davi, Farinaz Koushanfar, Andrew Paverd,
  Ahmad-Reza Sadeghi, and Gene Tsudik.
\newblock Invited: Things, trouble, trust: on building trust in {IoT} systems.
\newblock In {\em ACM/IEEE Design Automation Conference (DAC)}, 2016.

\bibitem{seda}
N~Asokan, Ferdinand Brasser, Ahmad Ibrahim, Ahmad-Reza Sadeghi, Matthias
  Schunter, Gene Tsudik, and Christian Wachsmann.
\newblock {SEDA}: Scalable embedded device attestation.
\newblock In {\em ACM SIGSAC Conference on Computer and Communications Security
  (CCS)}, 2015.

\bibitem{sabre-lite}
{Boundary Devices}.
\newblock i.mx6 arm development board.

\bibitem{brasser2015tytan}
Ferdinand Brasser, Brahim El~Mahjoub, Ahmad-Reza Sadeghi, Christian Wachsmann,
  and Patrick Koeberl.
\newblock {TyTAN}: tiny trust anchor for tiny devices.
\newblock In {\em ACM/IEEE Design Automation Conference (DAC)}, 2015.

\bibitem{brasser2016remote}
Ferdinand Brasser, Ahmad-Reza Sadeghi, and Gene Tsudik.
\newblock Remote attestation for low-end embedded devices: the prover's
  perspective.
\newblock In {\em ACM/IEEE Design Automation Conference (DAC)}, 2016.

\bibitem{lisa}
Xavier Carpent, Karim ElDefrawy, Norrathep Rattanavipanon, and Gene Tsudik.
\newblock Lightweigh swarm attestation: a tale of two {LISA}-s.
\newblock In {\em ACM Asia Conference on Computer and Communications Security
  (ASIACCS)}, 2017.

\bibitem{sha1}
D~Eastlake~3rd and Paul Jones.
\newblock Us secure hash algorithm 1 (sha1).
\newblock Technical report, 2001.

\bibitem{hydra}
Karim ElDefrawy, Norrathep Rattanavipanon, and Gene Tsudik.
\newblock Hydra: Hybrid design for remote attestation (using a formally
  verified microkernel).
\newblock {\em arXiv preprint arXiv:1703.02688}, 2017.

\bibitem{smart}
Karim Eldefrawy, Gene Tsudik, Aur{\'e}lien Francillon, and Daniele Perito.
\newblock {SMART}: Secure and minimal architecture for (establishing dynamic)
  root of trust.
\newblock In {\em Network and Distributed System Security Symposium (NDSS)},
  2012.

\bibitem{openmsp430}
Olivier Girard.
\newblock Openmsp430, 2009.

\bibitem{darpa}
Ahmad Ibrahim, Ahmad-Reza Sadeghi, Gene Tsudik, and Shaza Zeitouni.
\newblock {DARPA}: Device attestation resilient to physical attacks.
\newblock In {\em ACM Conference on Security and Privacy in Wireless and Mobile
  Networks (WiSec)}, 2016.

\bibitem{klein2009sel4}
Gerwin Klein, Kevin Elphinstone, Gernot Heiser, June Andronick, David Cock,
  Philip Derrin, Dhammika Elkaduwe, Kai Engelhardt, Rafal Kolanski, Michael
  Norrish, et~al.
\newblock sel4: Formal verification of an os kernel.
\newblock In {\em Proceedings of the ACM SIGOPS 22nd symposium on Operating
  systems principles}, pages 207--220. ACM, 2009.

\bibitem{trustlite}
Patrick Koeberl, Steffen Schulz, Ahmad-Reza Sadeghi, and Vijay Varadharajan.
\newblock {TrustLite}: A security architecture for tiny embedded devices.
\newblock In {\em ACM European Conference on Computer Systems (EuroSys)}, 2014.

\bibitem{sel4-libs}
{National ICT Australia and other contributors}.
\newblock sel4 libraries, 2014.

\bibitem{sel4-util-libs}
{National ICT Australia and other contributors}.
\newblock util\_libs, 2014.

\bibitem{ostrovsky1991withstand}
Rafail Ostrovsky and Moti Yung.
\newblock How to withstand mobile virus attacks.
\newblock In {\em Proceedings of the tenth annual ACM symposium on Principles
  of distributed computing}, pages 51--59. ACM, 1991.

\bibitem{blake2}
MJ~Saarinen and JP~Aumasson.
\newblock The blake2 cryptographic hash and message authentication code (mac).
\newblock 2015.

\bibitem{sana}
Ahmad-Reza Sadeghi, Matthias Schunter, Ahmad Ibrahim, Mauro Conti, and Gregory
  Neven.
\newblock {SANA}: Secure and scalable aggregate network attestation.
\newblock In {\em ACM Conference on Computer and Communications Security
  (CCS)}, 2016.

\bibitem{SCHELLEKENS200813}
Dries Schellekens, Brecht Wyseur, and Bart Preneel.
\newblock Remote attestation on legacy operating systems with trusted platform
  modules.
\newblock {\em Science of Computer Programming}, 74(1):13 -- 22, 2008.

\bibitem{Seshadri:2006:SSC:1161289.1161306}
Arvind Seshadri, Mark Luk, Adrian Perrig, Leendert van Doorn, and Pradeep
  Khosla.
\newblock Scuba: Secure code update by attestation in sensor networks.
\newblock In {\em ACM Workshop on Wireless Security (WiSe)}, 2006.

\bibitem{seshadri2004swatt}
Arvind Seshadri, Adrian Perrig, Leendert Van~Doorn, and Pradeep Khosla.
\newblock {SWATT}: Software-based attestation for embedded devices.
\newblock In {\em IEEE Symposium on Research in Security and Privacy (S\&P)},
  2004.

\bibitem{sha256}
Secure~Hash Standard.
\newblock Fips pub 180-2.
\newblock {\em National Institute of Standards and Technology}, 2002.

\bibitem{sha1-atk}
Marc Stevens, Elie Bursztein, Pierre Karpman, Ange Albertini, and Yarik Markov.
\newblock The first collision for full sha-1.
\newblock {\em URL: https://shattered. it/static/shattered. pdf}, 2017.

\bibitem{Stumpf2006}
Frederic Stumpf, Omid Tafreschi, Patrick R\"oder, and Claudia Eckert.
\newblock A robust integrity reporting protocol for remote attestation.
\newblock In {\em Workshop on Advances in Trusted Computing (WATC)}, 2006.

\bibitem{msp-gcc}
{Texas Instruments}.
\newblock Msp430-gcc-opensource gcc - open source compiler for msp
  microcontrollers, 2017.

\bibitem{stuxnet}
Jaikumar Vijayan.
\newblock Stuxnet renews power grid security concerns, june 2010.

\bibitem{xilinx2013}
ISE Xilinx.
\newblock Design suite, 2013.

\end{thebibliography}

\end{document}